\documentclass[prd,superscriptaddress,onecolumn,showkeys]{revtex4}
\usepackage{textcomp}
\usepackage{eurosym}
\usepackage{amsfonts}
\usepackage{array}
\usepackage{amsthm}
\usepackage{bm}
\usepackage{palatino}
\usepackage[colorinlistoftodos]{todonotes}
\usepackage{mathpazo}
\usepackage{supertabular}
\usepackage{subfig}
\usepackage{amssymb}
\usepackage{eurosym}
\usepackage{amsmath}
\usepackage{epsfig}
\usepackage{float}
\usepackage{graphics}
\usepackage{changes}
\usepackage{color}
\usepackage{graphicx}
\usepackage[colorlinks=true,
            linkcolor=blue,
           urlcolor=black,
           citecolor=black]{hyperref}

\def\be{\begin{equation}}
\def\ee{\end{equation}}
\def\beq{\begin{eqnarray}}
\def\eeq{\end{eqnarray}}

\def\bes{\begin{eqnarray}}
\def\ees{\end{eqnarray}}

\begin{document}
\title{Evaluation of physical properties of Kiselev like AdS spacetime in the context of $f(R,~T)$ gravity under the impact of quantum gravity}

\author{Riasat Ali}
\email{riasatyasin@gmail.com}
\affiliation{Department of Mathematics, Shanghai University and Newtouch Center for Mathematics of Shanghai University, Shanghai-200444, People's Republic of China}

\author{Xia Tiecheng}
\email{xiatc@shu.edu.cn}
\affiliation{Department of Mathematics, Shanghai University and Newtouch Center for Mathematics of Shanghai University, Shanghai-200444, People's Republic of China}

\author{~Rimsha Babar}
\email{rimsha.babar10@gmail.com}
\affiliation{Department of Mathematics, GC
University Faisalabad Layyah Campus, Layyah-31200, Pakistan}

\date{\today}

\begin{abstract}
The developments of the $f(R,~T)$ gravity theory, which is a logical expansion of general relativity according to Einstein, inspire us to examine this theory in greater detail and take up our study to obtain a modification of the Kiselev-like AdS black holes scenario. In this study, we employ the semi-classical Hamilton-Jacobi procedure to investigate the Hawking temperature $(T_{H})$ for 4-dimensional Kiselev-like AdS BHs in the context of $f(R,~T)$ gravity. We derive the temperature using a standard formula for Kiselev-like AdS BHs in the context of $f(R,~T)$ gravity. The relativistic field equation in the context of the generalized uncertainty principle (GUP), the semi-classical Hamilton-Jacobi procedure, and the WKB strategy are used to tunnel boson particles into the horizon under the effect of quantum gravity as well as Hawking temperature and also use this temperature to compute entropy corrections. Further, we study the $f(R,~T)$ gravity ($\zeta$), quantum gravity ($\alpha$), and particle kinetic energy ($\Xi$) effect on both temperature and entropy.
\end{abstract}

\keywords{Kiselev like AdS BHs; $f(R,~T)$ gravity; Semi-classical Hamilton-Jacobi procedure; WKB strategy; Tunneling radiation; Entropy corrections}

\date{\today}

\maketitle

\section{Introduction}
Albert Einstein's general relativity (GR) theory, which predicted black holes (BHs) in $1915$ \cite{A1, A2}, is still considered one of the world's most significant background ideas. According to GR, a BH is an area of space where the gravitational field is so strong that no matter can escape, including light. Black holes were considered to absorb anything and discharge nothing at all \cite{A3}; the quantum mechanics theory has fundamentally changed the Stephen Hawking concept. In $1974$, Hawking theoretically suggested that BHs have a specific temperature identified as the Hawking temperature ($T_H$) \cite{A4, A5}. Hawking’s calculations indicate that all BHs emit thermal energy through Hawking radiation. However, this effect becomes more pronounced for smaller, quantum-sized BHs due to higher temperatures, making the thermal energy emission particularly significant in such cases. Understanding quantum impacts in the context of classical general relativity was crucial to analyzing quantum gravity theories. One example of a quantum or GUP parameter \cite{F2} was reduced BH evaporation \cite{A6} by applying the Hamilton-Jacobi ansatz in which the modified Lagrangian equation and WKB approximation were applied. The radiation process was studied to combine classical mechanics with the gravitational theory of BH thermodynamics \cite{A7}. The literature has proposed several techniques for analyzing the Hawking radiation phenomenon. Numerous studies on the well-known BHs have focused on these radiations \cite{A8, A9, A10, A11, A12, A13, A14, A15, A16, A17, A18, A19, A20}. A semiclassical approach that depends on a particle flowing across the exterior horizon quantum tunneling of the BH from interior to exterior can be used to study these radiations \cite{A22, A23}. The null geodesic \cite{A24, A25, A26} and Hamilton–Jacobi methodology \cite{A27} techniques serve as the foundation for the tunneling strategy. Using the classically forbidden path of a particle, these approaches derive the imaginary component of the classical action from the exterior horizon.

An important development in our knowledge of quantum gravity is the GUP, which extends the classical Heisenberg Uncertainty Principle to include minimal length scales predicted by several quantum gravity frameworks. Critical to the resolution of spacetime singularities, modifications to BH thermodynamics and predictions of remnants at the end of BH evaporation are addressed by GUP by modifying the uncertainty relations. These contributions offer a strong foundation for investigating how quantum mechanics and gravity interact, with consequences for high-energy instances and Planck-scale physics. The lack of a broadly recognised formulation and dependency on free parameters create theoretical ambiguity, while experimental confirmation remains difficult due to the small scale of quantum gravitational effects without GUP parameter limitations. Despite GUP being a strong phenomenological instrument that provides a unifying perception across many quantum gravity theories while laying the path for future research into fundamental physics. Its relevance to BHs, more generalised thermodynamics and quantum corrections makes it.

The GUP relations are crucial for defining the nature of BHs, even though the quantum effects can be viewed as significant effects close to an event horizon of a BH. The quantum tunneling approach has been used to study the quantum effects, including GUP relations with thermodynamical properties for different spins of BHs \cite{A27a, A28, A29, A30, A31, A32, A33, A34, A35, A36, A37}. It is essential to note that quantum corrections significantly impact the Hawking temperature evolution rate during BH evaporation. While some theoretical models suggest that a BH may stop radiating at its lowest mass, leaving behind a remnant, this is still a hypothetical idea with no solid empirical proof. Furthermore, classical BH metrics do not automatically account for quantum phenomena; therefore, introducing quantum corrections necessitates a thorough analysis of quantum-corrected metrics and their consequences. Effective field theory (EFT) techniques have allowed for the recent determination of the quantum corrected metric of Schwarzschild BHs \cite{A38}, and they have also significantly advanced quantum gravity \cite{A39, A40, A41, A42, A43, A44, A45}. The EFT techniques can be used at energies lower than the Planck mass and allow model-independent quantum gravity calculations, even if analyzing a consistent quantum gravity at all energy scales remains very interesting. Gecim and Sucu investigated the Hawking temperature of the spin-1 particle by the tunneling approach, which was critical of the BH spacetime \cite{A45a} and GUP framework \cite{A45b, A45c}. The GUP framework showed that temperature was influenced not only by BH characteristics but also by tunneled particles. The expressions of their vector particles were based on the relativistic quantum spin-1 particle and the temperature of BHs of the vector bosons described by the modified equation.The modified equations have been applied \cite{A45d} to calculate the GUP corrections to the temperature of the Euler–Heisenberg-adS BH. The main features of the BH and its conventional thermodynamics were reviewed, and the usual thermodynamics in certain particle physical quantities were also compared with the improved model. For the zitterbewegung particles, the efficiency due to the GUP impact was greater than the standard temperature when attractive Coulomb interaction has been considered in \cite{A45e}.

A metric that reflects the structure of a BH with a massive object was created by mapping an emergent-gravity metric that incorporates $k$-essence scalar fields with a Lagrangian of Born-Infeld \cite{A45f} and Lagrangian of Dirac–Born–Infeld \cite{A45g} and the metric's Hawking temperature have been calculated. The application of the Hamilton–Jacobi strategy for tunnelling with both rotation parameter BH and non-rotation BH in various non-singular coordinate frames not only validates this quantum radiation from BHs with a frame of dark energy but also introduces the novel phenomenon of absorption into white holes by quantum mechanical tunnelling \cite{A45h}. Mitra studied \cite{A45i} radiation tunnels through BH horizons, which leads to a temperature double the usual value.

In thermal phenomena, the entropy of BHs can be affected by quantum corrections. In contrast to thermal effects, which are related to the classical thermodynamic behavior of BH as defined by the Bekenstein-Hawking entropy, quantum corrections consider changes resulting from quantum gravity theories. Both contributions are considered within a single framework in a more comprehensive strategy. Highly sophisticated techniques, including Hamiltonian split functions \cite{A46}, near-horizon symmetries, and loop quantum gravity \cite{A47}, have been employed to compute the quantum corrections to BH entropy. To study the logarithmic corrections to BH entropy, a canonical entropy that permits thermal fluctuations is more suitable than micro-canonical entropy \cite{A48}. From a thermodynamic perspective, nothing can cross the horizon under an isolated horizon boundary, and thermal fluctuations are forbidden to the horizon areas in this frame. It is demonstrated that the Schwarzschild BH will be in thermal equilibrium since it has a specific heat, suggesting that its thermodynamic properties and entropy are precisely defined. By immersing the BH in an iso-thermal bath, it can be contained within a finite-radius cavity, and its surroundings will eventually reach thermal equilibrium. A regularly spaced area spectral \cite{A49} has been identified in the literature by Bekenstein and other authors \cite{E13}. It is noted in \cite{A50} that the local temperature for a local observer at rest is dependent on the observer's position. A canonical boundary condition leads us to conclude that either two solutions or none may be found. A higher mass solution has a higher specific heat, whereas a smaller one has a lower specific heat \cite{A50, A51}. The modifications to the Schwarzschild BH with the more mass solution based on a regularly spaced area spectrum phenomenon have been discussed. To attain thermal equilibrium, the surround of Schwarzschild BH in a limited cavity of finite size has been analyzed in \cite{A52}. The Schwarzschild BH was never in thermal equilibrium due to the logarithmic corrections of thermal fluctuations on the entropy.
Further, the researchers examined the relationship between other thermal characteristics and the modified entropy for issues of stability as well as the mechanics of several kinds of BHs according to their impact on correction factors\cite{F1, F3, F4, F5, F6, F7}. 

Determining tunneling radiation in outer horizon spacetime is significant using several well-established techniques. However, as we study the spacetimes in $f(R,~T)$ gravity theory, it becomes even more crucial to comprehend $T_H$. The BH has a standard $T_H$ raised by the general formula. According to our research, all of the $f(R,~T)$ contributions significantly contribute to $T_H$. In this study, we investigate the $T_H$ of a BH in four dimensions, which is also $f(R,~T)$ gravity theory.
Furthermore, we establish that the quantum gravity of this BH contributes to the $T_H$. This $T_H$ is comparable to a Kiselev-like AdS BHs $T_H$. Each particle expels the mass and energy from the Kiselev-like AdS BHs lost to Hawking radiation, which lowers mass and causes evaporation. Its radiation is emitted inside the speed uncertainty region. The procedure described the law of energy and momentum proposed by relativistic physics principles. This semi-classical approach corresponds to the derivation algorithm of $T_H$, which combines the uncertainty principle from quantum mechanics with the Kiselev-like AdS BHs radius from GR. The semi-classical approach is widely used in physics to describe phenomena that require different explanations. However, this approach can result in approximations with modified validity. Our analysis of particle generation depends on the properties of quantum fields near the horizon. The particle is radiated to properly ignore self-attraction, self-gravity, and back reaction.

This work proposes to investigate the Hawking radiation of significant vector particles from the Kiselev like AdS spacetime in the context of $f(R,~T)$ gravity. The Standard Model focuses significantly on vector particles (bosons of spin-1), which include the well-known $W^{\pm}$ bosons. The Lagrangian equation can represent the trajectory of the boson field when we take into account the immense bosons in space-time. In this scenario, we can explore the tunnelling process using the Hamilton-Jacobi ansatz and the WKB approximation to the Lagrangian solution in space-time, following a comparable approach as in \cite{F2}. The $f(R,~T)$ gravity metric massive boson ($W^{\pm}$) fields interact with the dust, radiation, phantom and quintessence fields, in which study their motion is more complex than in the Lagrangian equation. First, we use the Glashow-Weinberg-Salam model's Lagrangian to obtain the $W^{\pm}$ boson field equation. We use the Hamilton-Jacobi ansatz and WKB approximation to solve the equation that follows in Kiselev like AdS spacetime in the context of $f(R,~T)$ gravity. To set the coefficient matrix determinant to zero, then we can solve the linear function for the radial function. We analyse the tunnelling rate associated with the Hawking temperature of significant vector particles from the Kiselev like AdS spacetime in the context of $f(R,~T)$ gravity and derive the related entropy.

A comparison with current observational data demonstrates how this work may be relevant to astrophysical processes. Observations of BH applications using high-energy electromagnetic signals (e.g., Event Horizon Telescope) and gravitational wave detections (e.g., LIGO and Virgo) indirectly show Hawking-like radiation effects and BH thermodynamics. However, because of its small signal compared to the surrounding cosmic background, exact measurements of Hawking radiation are still tricky. This finding supports attempts to understand these observational data using theoretical models such as gravity and introducing quantum corrections via GUP. For example, the temperature profiles obtained in this work may be able to explain variations in the thermal spectra of compact objects or the behaviors of BH accretion discs. Future experimental designs that seek to identify quantum gravity evidence in BH radiation are also based on these theoretical assumptions.

In this article, we examine Kiselev-like AdS BHs in the context of $f(R,~T)$ gravity in Section \textbf{II} and also study the $T_H$ by a standard formula. Section \textbf{III} generates the tunneling radiation. Graphics are analyzed for the corrected $T_H$ physical stability and instability. In Sec. \textbf{IV}, describe corrected entropy. In section \textbf{V}, we finalize our conclusion.

\section{Kiselev-like AdS BHs solution in the context of $f(R,~T)$ gravity}
Kiselev determined \cite{F12} the precise solutions to Einstein's equations for the static, spherically symmetric fundamental state; it was surrounded by a charged BH. These solutions were based on the state parameter $\omega$ and the unique internal parameter selection in the energy-momentum tensor of quintessence. In accordance with the basic addition of their contributions in the total energy-momentum tensor, the linear expressions for the different matter terms were produced by the relevant conditions of additivity and linearity. It obtains accurate bounds on the known solutions for the unusual case of the de Sitter space or cosmological constant as well as for the electromagnetic static field, which implies the relativistic relationship between the pressure and energy density. To categorise the horizons, which are clearly visible in the static coordinates. In the framework of $f(R,~T)$ gravity, we want a static spherically symmetric AdS black hole solution provided by Kiselev. To do this, we start with the $4$-dimensional Einstein-Hilbert action for $f(R,~T)$ gravity \cite{F8, F9} as
\begin{equation}
\mathbb{S}=\frac{1}{16\pi}\int (f(R,~T)-2\Lambda)\sqrt{-g}d^{4}x+\int\sqrt{-g}d^{4}x\mathbb{L_{m}},\label{L1}
\end{equation}
where $R$, trace $T$, and $g$ represent the Ricci scalar, the energy-momentum tensor, and metric tensor $g_{\mu\nu}$ determinant, respectively. An anisotropic fluid \cite{F9, F9a} that is effectively coupled to Kiselev BHs is given by $\mathbb{L_{m}}$ as 
\begin{equation}
\mathbb{L_{m}}=-\frac{1}{3}(2p_{t}+p_{r}),\label{L2}
\end{equation}
and the negative cosmological constant $(\Lambda= \frac{-3}{l^{2}})$, where $l$ is the AdS length scale. We may now obtain equations of motion by varying the action (\ref{L1}) about the metric tensor $(g_{\mu\nu})$ \cite{F10, F11} as
\begin{equation}
f_{R}(R,~T)R_{\mu\nu}-\frac{1}{2}g_{\mu\nu}f(R,~T)+(g_{\mu\nu}\square-\nabla_{\mu}\nabla_{\nu})f_{R}(R,~T) -\frac{\Lambda}{2}g_{\mu\nu}=8\pi T_{\mu\nu}-T_{\mu\nu}f_{T}(R,~T)-\Theta_{\mu\nu}f_{T}(R,~T),\label{L3}
\end{equation}
with $f_{T}(R,~T)$ and $f_{R}(R,~T)$ stand for the corresponding derivatives of $f(R,~T)$ with regard to $T$ and $R$. Further $\Theta_{\mu\nu}$ can be defined as
\begin{equation}
\Theta_{\mu\nu}=g^{\alpha\beta}\frac{\partial T_{\alpha\beta}}{\partial g_{\mu\nu}}=-2T_{\mu\nu}+g_{\mu\nu}\mathbb{L}_{m}-2g^{\alpha\beta}\frac{\partial^{2}\mathbb{L}_{m}}{\partial g^{\mu\nu}g^{\alpha\beta}}.\label{L4}
\end{equation}
We take into consideration the following particular form of $f(R,~T)$ for our investigations \cite{FF10} as
\begin{equation}
f(R,~T)=R+2f(T),\label{L5}    
\end{equation}
which implies to
\begin{equation}
R_{\mu\nu}-\frac{g_{\mu\nu}R}{2}-g_{\mu\nu}\Lambda=8\pi T_{\mu\nu}-\frac{2df(T)}{dT}(T_{\mu\nu}+\Theta_{\mu\nu})+g_{\mu\nu}f(T).\label{L6}  
\end{equation}
The following line element can be used to characterize the static geometry in spherically symmetric spacetime that we are assuming as
\begin{equation}
ds^{2}=F(r)dt^{2}-\frac{1}{F(r)}dr^{2}-r^{2}d\Omega_{D-2}^{2}.\label{L7}
\end{equation}
Now, identify the form of the unknown function $F(r)$. The energy-momentum tensor components for Kiselev BHs are as shown in \cite{F12} as
\begin{eqnarray}
T^{r}_{r}&=&T^{t}_{t}=\rho(r),\\ \label{L8}
T^{\phi}_{\phi}&=&T^{\theta}_{\theta}=-\frac{\rho}{2}(1+3\omega), \label{L9}
\end{eqnarray}
where $\omega$ is the equation of the state parameter. The energy-momentum tensor of Kislev BHs \cite{F12} is comparable to that of an anisotropic fluid, which implies that
\begin{equation}
T_{\nu}^{\mu}=diag(p_{t} , -p_{r}, -p_{\theta}, -p_{\phi})\label{L10}
\end{equation}
with $p_{t}=-\frac{1}{2\rho}(1+3\omega)$ and $p_{r}=-\rho$. 
The tensor of energy-momentum in Eq. (11) can be derived from the generic form of an anisotropic fluid's tensor of energy-momentum \cite{F13} as
\begin{equation}
T_{\mu\nu}=-p_{t}g_{\mu\nu}+(\rho+p_{t})\mathcal{U}_{\nu}\mathcal{U}_{\mu}-(p_{t}-p_{r})\mathcal{N}_{\mu}\mathcal{N}_{\nu}, \label{L11}
\end{equation}
where $p_{r}(r)$, $\rho$, and $p_{t}(r)$ represent the radial of fluid pressure, energy density, and tangential, respectively.
The physical object, $\mathcal{U}^{\mu}=(F(r),~0,~0,~0)$ and $\mathcal{N}^{\mu}=(0,~\frac{1}{F(r)},~0,~0)$ are 4-velocity and vector of radial unit, respectively, that satisfies with the conditions $\mathcal{U}_{\nu}\mathcal{U}^{\nu}=1,~~\mathcal{N}_{\nu}\mathcal{N}^{\nu}=-1 $ and $\mathcal{U}_{\nu}\mathcal{N}^{\nu}=0$. By employing Eq. (\ref{L2}) to Eq. (\ref{L4}), we have the

\begin{equation}
\Theta_{\mu\nu}=-2T_{\mu\nu}-\frac{g_{\mu\nu}}{3}(2p_{t}+p_{r}).\label{L12}
\end{equation}
Considering the expression $f(T)=\zeta T$ and the application of equations (\ref{L6}), (\ref{L7}), and (\ref{L12}), to derived the unknown function \cite{F9} as
\begin{equation}
F(r)=1-\frac{2\mathcal{M}}{r}+\frac{r^{2}}{l^{2}}+\frac{k}{r^{\frac{8(\pi+\omega\zeta+3\pi\omega)}{8\pi+3\zeta-\omega\zeta}}},\label{L13}
\end{equation}
where $\mathcal{M}$ is the BH mass, $k$ is the integration constant, and $\zeta$ is the $f(R,~T)$ gravity parameter. The well-known Kiselev-AdS BH solution in GR can be obtained \cite{F14} by assuming $\zeta\rightarrow 0$, we get 
\begin{equation}
F(r)|_{\zeta\rightarrow 0}=1-\frac{2\mathcal{M}}{r}+\frac{r^{2}}{l^{2}}+\frac{k}{r^{(1+3\omega)}}.\label{L14}
\end{equation}
Now, we wish to consider the BH solution's horizon structure. The location of the BH horizon can be found by solving $F(r)=0$, as is widely known. Since we were unable to determine the horizon analytically in our situation, we used a numerical method to determine the horizons by visualizing $F(r)$ vs. radial coordinate $r$ in \cite{F9}, for various values of the $f(R,~T)$ gravity parameter $\zeta$ and the equation of state parameter $\omega$ to determine that the parameter $\zeta$ for a given $\omega$  determines the number of horizons that are conceivable. For example, $\omega=-\frac{4}{3}$ (phantom field), $\omega = -\frac{2}{3}$ (quintessence field), $\omega= \frac{1}{3}$ (radiation field) and $\omega=0$ (dust field) are among the specific circumstances that we take into consideration. In \cite{F9}, we have two horizons (Cauchy and event) for dust, radiation, and the quintessence field. By solving $F(r_{+})=0$, one can determine the BH mass as
\begin{equation}
\mathcal{M}=\frac{r_{+}}{2}\Big(1+kr_{+}^{-\frac{8(\pi+\zeta \omega+3\pi\omega)}{8\pi+3\zeta-\omega\zeta}}+\frac{r^{2}_{+}}{l^{2}}\Big), \label{L15}
\end{equation}
with $r_{+}$ represent the event horizon.

The Hawking temperature for Kiselev-like AdS spacetime in the context of $f(R,~T)$ gravity can be obtained by applying the standard formula throughout the scientific literature: $T_{H}=\frac{\textsc{K}(r_{+})}{2\pi}$ under the influence of the surface gravity $(\textsc{K}(r_{+}))$ in this manner as
\begin{equation}
T_{H}=\frac{1}{4\pi}\left[\frac{{1}}{r_{+}}+\frac{3r_{+}}{l^{2}}+\frac{3k(\zeta-8\pi\omega-3\omega)r_{+}^{\frac{8\pi(2+3\omega)+\zeta(3+7\omega)}{-8\pi-3\zeta+\omega\zeta}}}{8\pi+3\zeta-\omega\zeta}\right].
\end{equation}
The $T_{H}$ is dependent on the $\omega$ parameter, the $f(R,~T)$ gravity $(\zeta)$ parameter, the cosmological constant. When $\zeta=0$, we get Hawking temperature Kiselev-like AdS spacetime. If $r_{+}=2\mathcal{M}$, the Schwarzschild spacetime Hawking temperature is similar to the first term of our Hawking temperature.

The analysis below explores temperature graphs that illustrate the link between temperature $T_{H}$ and horizon radius $r_{+}$ under different values of the parameter $(\zeta)$. Each graph illustrates the system's thermal behavior as $r_{+}$ grows, including crucial observations on how $(\zeta)$ affects the temperature profile.
\begin{figure}[H]
\centering
\includegraphics[width=6cm,height=6cm]{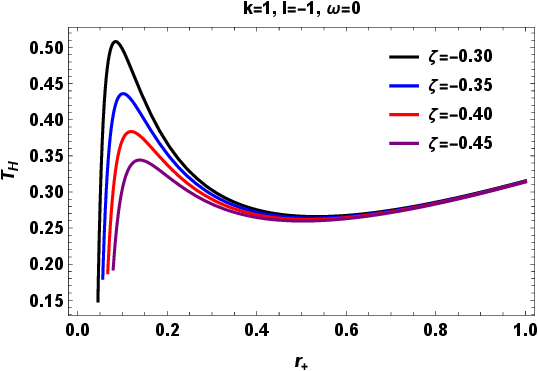}\includegraphics[width=6cm,height=6cm]{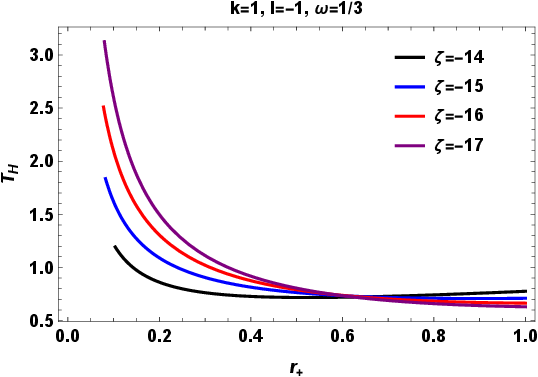}
\caption{Temperature $T_H$ via horizon radius $r_+$ for dust field ($\omega=0$) and radiation field ($\omega=1/3$) under the influence of $f(R,~T)$ gravity parameter $(\zeta)$.}\label{t11}
\end{figure}
In the left graph of Fig. \ref{t11}, \( T_H \) first rises with \( r_+ \) until it peaks, at which point it progressively falls. With varying \( \zeta \) values ($0.30,~0.35,~ 0.40,~0.45)$, the peaks fluctuate somewhat, and the temperature decreases more slowly after the peak than it did before. The temperature peak suggests the existence of a maximum stable temperature that varies when \( \zeta \) fluctuates. This peak may indicate a critical point or phase transition in the system, beyond which cooling results from a change in the thermal characteristics. This behavior suggests that \( \zeta \) influences thermal stability and critical points, with lower \( \zeta \) values allowing higher temperatures at smaller radii. The curves shift downward as \( \zeta \) values increase, showing that higher \( \zeta \) values lower the maximum achievable temperature.

In the right plot of Fig. \ref{t11}, for various values of \( \zeta \) $(14,~15,~16,~17)$, the graph indicates a reduction in temperature \( T_h \) as the horizon radius \( r_+ \) grows.
As \( r_+ \) rises, the temperature drops dramatically and stabilizes close to zero for larger values of \( r_+ \). As the horizon radius increases, this pattern points to a potential approach to a minimum temperature, suggesting that the system could achieve thermal equilibrium at more enormous radii.
There is a clear trend where higher \( \zeta \) values result in greater temperatures at any given \( r_+ \), even though the distance between curves for various \( \zeta \) values is minimal. According to this finding, \( \zeta \) could be a control parameter that affects how quickly the temperature decays.

From Fig. \ref{t11}, we conclude that the thermal response is always influenced by the parameter \( \zeta \), and its value determines whether the temperature rises or falls. The connection between peak temperature and \( \zeta \) is inverse. While the right graph shows a critical threshold behavior, in which the temperature increases to a maximum before stabilizing at a lower level, the left graph implies that the temperature stabilizes at greater radii. A physical system where $\zeta$ may represent an external field or interaction strength that affects the thermal equilibrium states is probably modeled by the temperature behavior dependent on $r_+$ and $\zeta$.

\begin{figure}[H]
\centering
\includegraphics[width=6cm,height=6cm]{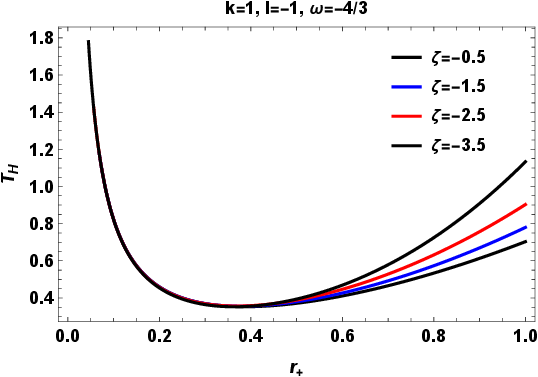}\includegraphics[width=6cm,height=6cm]{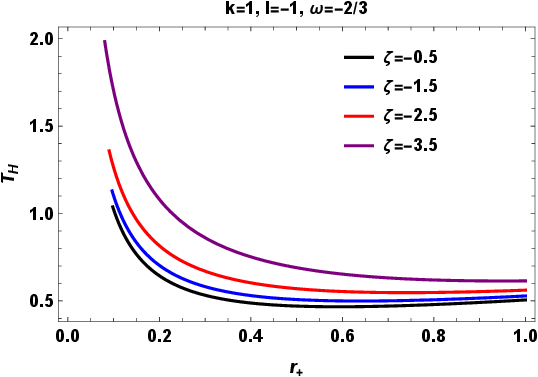}
\caption{Temperature $T_H$ via horizon radius $r_+$ for phantom field ($\omega=-4/3$) and quintessence field ($\omega=-2/3$) under the influence of $f(R,~T)$ gravity parameter $(\zeta)$.}\label{t22}
\end{figure}
The following analysis examines the behavior of temperature \( T_h \) at the horizon \( r_+ \) for various values of \( \zeta \) in a system defined by parameters \( k = 1 \), \( l = -1 \) under the influence of phantom ($\omega=-4/3$) and quintessence field ($\omega=-2/3$). 

Curves for various values of \( \zeta \) (ranging from $0.5$ to $3.5$) are shown in both graphs with different colors.
A noticeable change in the temperature behavior occurs when \( \zeta \) rises. The temperature curves exhibit varying slopes for more significant \( \zeta \), suggesting that temperature depends on \( \zeta \) close to the horizon. The plots show that the parameters \( \zeta \) and \( \omega \) influence the non-trivial connection between temperature and horizon radius.

In the left plot of Fig. \ref{t22}, the temperature begins low for the $\omega=-4/3$ case and rises gradually as \( r_+ \) grows. There is a dramatic beginning fall at tiny \( r_+ \) values, especially for smaller \( \zeta \) values. 

In the right plot of Fig. \ref{t22}, the temperature values are often greater for \( \omega = -2/3 \) compared to the $\omega=-4/3$ scenario. The initial decrease is steeper for lower \( \zeta \), and the curves look more dispersed than in the $\omega=-4/3$ graph. This suggests that temperature is more strongly dependent on \( \zeta \) for this value of \( \omega \). At \( \omega = -2/3 \), the higher dependency on \( \zeta \) may indicate a threshold effect, in which specific parameter combinations result in a fundamentally different thermal response.

From Fig. \ref{t22}, we conclude that increasing \( \zeta \) often results in greater temperatures at bigger \( r_+ \) values for both values of \( \omega \), indicating that \( \zeta \) affects the system's thermal stability or rate of temperature growth.
The behavior of the temperature at low \( r_+ \) implies that the system is more sensitive to parameter changes for tiny BHs or horizons, which may indicate phase transitions or critical events related to horizon formation.

\section{Modified field equations and tunneling under the impact of quantum gravity}

A minimal measurable length is a characteristic shared by several theories of quantum gravity, including non-commutative geometry, loop quantum gravity, and string theory \cite{F15, F16, F17, F18}. It is feasible to realize this minimal length using the GUP \cite{F19, F20, F21}. In one-dimensional quantum mechanics, an effective model of the GUP that comprises the main concept of massive additional dimensions was provided \cite{F22} as
\begin{eqnarray}
\ell_{F}k_i(p)&=&\tanh\Big(\frac{p}{M_{F}}\Big),\\
\ell_{F}\omega_{1}(E)&=&\tanh\Big(\frac{E}{M_{F}}\Big),
\end{eqnarray}
with $\ell_{F}$ and $M_{F}$ stand for the higher-dimensional minimal length and Planck mass, respectively, and the wave vector $k_i$ and frequency $\omega_{1}$ are the operators of the changes in space and time. Then $\ell_{F}M_{F}=\textit{h}$ is satisfied by $\ell_{F}$ and $M_{F}$. Position representation $\hat{x}=x$ quantisation results in
\begin{equation}
\omega_{1}=i\partial_{t},~~~k_i=-i\partial_{x}.
\end{equation}
Consequently, taking into account the order of $(\frac{p}{M_{F}})^{3}$, the low energy limit $p<<M_{F}$ provides to

\begin{eqnarray}   
E&\simeq &i\textit{h}\partial_{t}\Big(1-\alpha\textit{h}\partial^{2}_{t}\Big),\label{GU1}\\
p&\simeq &- i\textit{h}\partial_{x}\Big(1-\alpha\textit{h}\partial^{2}_{x}\Big),\label{GU2}
\end{eqnarray}
with $\alpha=\frac{1}{3M^{2}_{F}}$. So, the modified commutation relation is denoted by
\begin{equation}
[x,p]=i\textit{h}\big(1+\alpha p^{2}\Big),\label{GU3}
\end{equation}
and the relation of generalized uncertainty is 
\begin{equation}
\Delta x\Delta p\geq \frac{\textit{h}}{2}\Big[1+\alpha(p^{2})\Big]. \label{GU4}
\end{equation}
It is evident from Eqs. (\ref{GU3}) and (\ref{GU4}) that as the particle's momentum extends, so does GUP's divergence from the principle of Heisenberg uncertainty. It should be noted that the specific scenario examined in this paper, which is the low energy limit $p<<M_{F}$, is the only one for which Eqs. (\ref{GU1})–(\ref{GU4}) apply. Experiments such as those by \cite{F23, F24} that aim to test the uncertainty principle should confine the parameter $\alpha$ in the low energy region. Previous investigations have identified further generalized uncertainty relationships. By employing the Hamilton–Jacobi approach, which considers the minimal length effect via Eqs. (\ref{GU1}) and (\ref{GU2}), we examine how massive spin-1 particles tunnel across BH horizons. Our models reveal that the quantum gravity modification is related to the BH's mass and the mass and angular momentum of the emitted boson particles. Furthermore, the quantum gravity correction explicitly delays the rise in temperature during the BH evaporation process, which means that the quantum correction will balance the traditional tendency for a temperature increase at some point during the evaporation, which results in the formation of remnants.

Using the Lagrangian equation, we examined particle tunneling beyond a spacetime horizon. The real and imaginary calculation indicates that the corrected Hawking temperature for tunneling a vector particle depends on the particle's mass. The corrected Hawking temperature of spacetime is discovered to depend on both the mass and kinetic energy of the expelled particle if a small term in the roots of a quadratic equation with physical meaning is ignored. According to the actual calculation, the corrected Hawking temperature for the vector tunneling across the space-time mass depends on the $f(R,~T)$ gravity, surround field, and cosmological constant of the expelled particle. Within the paradigm of GUP, we begin with the kinetic term of the uncharged vector boson field in flat spacetime, $\frac{1}{2}\Tilde{B}_{\mu\nu}\Tilde{B}^{\mu\nu}$, with the modified field strength tensor \cite{F2} is provided by
\begin{equation}
\Tilde{B}_{\mu\nu}=\Big(1-\alpha \textit{h}^{2}\partial^{2}_{\mu}\Big)\partial_{\mu}B_{\nu}-\Big(1-\alpha \textit{h}^{2}\partial^{2}_{\nu}\Big)\partial_{\nu}B_{\mu}.
\end{equation}
Considering the gauge principle, the additional derivatives also act on the local unitary transformation operator $U(x)$, so covariant derivatives must also replace them \cite{F25}. It should be noted that additional derivative terms exist. Next, we generalize this to the case of a vector field $(\textbf{W}^{\pm})$ in BH spacetime.
\begin{eqnarray}
\Big(1-\alpha \textit{h}^{2}\partial^{2}_{0}\Big)\partial_{0}&\rightarrow & \Big(1+\alpha \textit{h}^{2}g^{00}\nabla_{0}^{\pm 2}\Big)\nabla_{0}^{\pm},\label{GU5}\\
\Big(1-\alpha \textit{h}^{2}\partial^{2}_{i}\Big)\partial_{i}&\rightarrow & \Big(1-\alpha \textit{h}^{2}g^{ii}\nabla_{i}^{\pm 2}\Big)\nabla_{i}^{\pm},\label{GU6}
\end{eqnarray}
with $\nabla_{\mu}$ representing the geometrical covariant derivative. The $g^{00}$ and $g^{ii}$ always have distinct signs, which explains the discrepancy in signs of the $O(\alpha)$ component in Eqs. (\ref{GU5}) and (\ref{GU6}). By specifying
\begin{eqnarray}
\nabla_{o}^{\pm}=\Big(1+\alpha \textit{h}^{2}g^{00}\nabla_{0}^{\pm 2}\Big)\nabla_{0}^{\pm}~~~\textit{and}~~~\nabla_{i}^{\pm}=\Big(1-\alpha \textit{h}^{2}g^{ii}\nabla_{i}^{\pm 2}\Big)\nabla_{i}^{\pm}.\nonumber
\end{eqnarray}
In the $\textbf{W}$-boson field \cite{F2}, the corrected Lagrangian for GUP is provided as
\begin{equation}
\mathcal{L}^{GUP}=-\frac{1}{2}\Big(\nabla_{\mu}^{+}\textbf{W}_{\nu}^{+}-\nabla_{\nu}^{+}\textbf{W}_{\mu}^{+}\Big)\Big(\nabla^{-\mu}\textbf{W}^{-\nu}-\nabla^{-\nu}\textbf{W}^{-\mu}\Big)-\frac{m_{\textbf{W}}}{\textit{h}}\textbf{W}_{\mu}^{+}\textbf{W}^{-\mu}.
\end{equation}
Hence, the appropriate generalized action that should be taken as
\begin{equation}
\mathbb{S}^{GUP}=\int d^{4}x\sqrt{-g} \mathcal{L}^{GUP}\big(\textbf{W}_{\mu}^{\pm},~\partial_{\mu}\textbf{W}_{\nu}^{\pm},~\partial_{\rho}\partial_{\mu}\textbf{W}_{\nu}^{\pm},~\partial_{\rho}\partial_{\mu}\partial_{\lambda}\textbf{W}_{\nu}^{\pm}).\label{GU7}
\end{equation}
The physical significance of the GUP component of the Lagrangian equation is considered. The GUP parameter is the field equation without a singularity extended as a Lagrangian field equation. We use the boson field (vector field) in the Lagrangian equation of action to examine the boson radiation phenomenon, such as
\begin{equation}
\partial_{\mu}(\sqrt{-g}B^{\nu\mu})+\sqrt{-g}\frac{m^2}{\textit{h}^2}B^{\nu}+
\alpha\textit{h}^{2}\partial_{0}\partial_{0}\partial_{0}(\sqrt{-g}g^{00}B^{0\nu})
-\alpha \textit{h}^{2}\partial_{i}\partial_{i}\partial_{i}(\sqrt{-g}g^{ii}B^{i\nu})=0,\label{GU8}
\end{equation}
In this case, the coefficient matrix determinant, boson particle mass, and anti-symmetric tensor are represented by the $g$, $B^{\nu\mu}$, and $m$. The anti-symmetric tensor can be defined as
\begin{equation}
B_{\nu\mu}=(1-\alpha{\textit{h}^2\partial_{\nu}^2})\partial_{\nu}B_{\mu}-
(1-\alpha{\textit{h}^2\partial_{\mu}^2})\partial_{\mu}B_{\nu}\nonumber,
\end{equation}
with $\alpha$ and $\textit{h}$ are the GUP parameter and Plank's constant, respectively.
It is conceivable to compute the components of $B^{\mu}$ and $B^{\mu\nu}$ as

\begin{eqnarray}
&&B^{0}=-\frac{1}{F(r)}B_{0},~~~B^{1}=F(r)B_{1},~~~B^{2}=\frac{1}{r^{2}}B_{2},~~~
B^{3}=\frac{1}{r^{2}\sin^2\theta}B_{3},~~~B^{01}=-B_{01},B^{02}=-\frac{1}{F(r)r^{2}}B_{02},\nonumber\\
&&B^{03}=-\frac{1}{F(r)r^{2}\sin^2\theta}B_{03},~~~
B^{12}=\frac{F(r)}{r^{2}}B_{12},~~~B^{13}=\frac{F(r)}{r^{2}\sin^2\theta}B_{13},~~~
B^{23}=\frac{1}{r^{4}\sin^2\theta}B_{23}.\nonumber
\end{eqnarray}
A significant field of the theory of quantum gravity effects is studying BHs, and the GUP has been used in numerous BH physics experiments. Within the context of GUP, the thermodynamics of BH have been studied \cite{F26}. Nozari and Mehdipour investigated the improved tunneling rate of a Schwarzschild BH \cite{F27} by integrating the GUP with the tunneling algorithm. In \cite{F26}, the corrected Hawking temperatures were computed for different spacetime forms, and the GUP-deformed Hamilton–Jacobi equation for bosons in curved spacetime was introduced. This property naturally leads to a recovered mass in BH evaporation; however, it was discovered by analyzing the tunneling of bosons that the effects of quantum gravity slowed down the increase in Hawking temperatures. The WKB approach is stated by \cite{F28}
\begin{equation}
B_{\nu}=d_{\nu}\exp\left[\frac{i}{\textit{h}}D_{0}(t,r,\theta,\phi)+
\Sigma \textit{h}^{n}D_{n}(t,r,\theta,\phi)\right].
\end{equation}
The constant term is denoted by $d_{\nu}$, while $(D_{0},~D_{n})$ represents arbitrary functions. By ignoring higher orders in the corrected Lagrangian for GUP (\ref{GU8}) and only considering the term $\textit{h}$ in the WKB approximate for the $1^{st}$ order, we obtain the equation system given as 

\begin{eqnarray}
&&F(r)\left[d_{1}(\partial_{0}D_{0})(\partial_{1}D_{0})+\alpha d_{1}
(\partial_{0}D_{0})^{3}(\partial_{1}D_{0})-d_{0}(\partial_{1}D_{0})^{2}
-\alpha d_{0}(\partial_{1}D_{0})^4\right]+\nonumber\\
&&\frac{1}{r^{2}}\left[d_{2}(\partial_{0}D_{0})(\partial_{2}D_{0})+\alpha d_{2}
(\partial_{0}D_{0})^3(\partial_{2}D_{0})-d_{0}(\partial_{2}D_{0})^2-\alpha d_{0}
(\partial_{2}D_{0})^4\right]+\nonumber\\
&&\frac{1}{r^{2}\sin^2\theta}[d_{3}(\partial_{0}D_{0})(\partial_{3}D_{0})
+\alpha d_{3}(\partial_{0}D_{0})^3(\partial_{3}D_{0})
-d_{0}(\partial_{3}D_{0})^2-\alpha d_{0}(\partial_{3}D_{0})^4]-d_{0}m^2=0,\label{aa}\\
&&-\frac{1}{F(r)}\left[d_{0}(\partial_{0}D_{0})(\partial_{1}D_{0})+\alpha d_{0}
(\partial_{0}D_{0})(\partial_{1}D_{0})^3-d_{1}(\partial_{0}D_{0})^{2}
-\alpha d_{1}(\partial_{0}D_{0})^{4}\right]+\nonumber\\
&&\frac{1}{r^{2}}\left[d_{2}(\partial_{1}D_{0})(\partial_{2}D_{0})+\alpha d_{2}
(\partial_{1}D_{0})^3(\partial_{2}D_{0})-d_{1}(\partial_{2}D_{0})^{2}-\alpha
d_{1}(\partial_{2}D_{0})^{4}\right]+\nonumber\\&&
\frac{1}{r^{2}\sin^2\theta}\left[d_{3}(\partial_{1}D_{0})(\partial_{3}D_{0})+\alpha d_{3}
(\partial_{1}D_{0})^3(\partial_{3}D_{0})
-d_{1}(\partial_{3}D_{0})^{2}-\alpha d_{1}(\partial_{3}D_{0})^{4}\right]-d_{1}m^2=0,\\
&&-{\frac{1}{F(r)}}\left[d_{0}(\partial_{0}D_{0})(\partial_{2}D_{0})+\alpha d_{0}
(\partial_{0}D_{0})(\partial_{2}D_{0})^{3}-d_{2}(\partial_{0}D_{0})^{2}
-\alpha d_{2}(\partial_{0}D_{0})^4\right]\nonumber\\
&&+F(r)\left[d_{1}(\partial_{1}D_{0})(\partial_{2}D_{0})+\alpha d_{1}
(\partial_{1}D_{0})(\partial_{2}D_{0})^{3}
-d_{2}(\partial_{1}D_{0})^{2}-\alpha d_{2}(\partial_{1}D_{0})^4\right]+\nonumber\\
&&\frac{1}{r^{2}\sin^2\theta}\left[d_{3}(\partial_{2}D_{0})(\partial_{3}D_{0})+\alpha d_{3}
(\partial_{2}D_{0})^{3}(\partial_{3}D_{0})-d_{2}(\partial_{3}D_{0})^{2}-\alpha d_{2}(\partial_{3}D_{0})^4\right]+m^2 d_{2}=0,\\
&&-{\frac{1}{F(r)}}\left[d_{0}(\partial_{0}D_{0})(\partial_{3}D_{0})+\alpha d_{0}
(\partial_{0}D_{0})(\partial_{3}D_{0})^{3}-d_{0}(\partial_{3}D_{0})^{2}-
d_{0}(\partial_{3}D_{0})^{4}\right]-\nonumber\\
&&F(r)\left[d_{1}(\partial_{1}D_{0})(\partial_{3}D_{0})+\alpha d_{1}
(\partial_{1}D_{0})(\partial_{3}D_{0})^{3}-d_{3}(\partial_{1}D_{0})^{2}
-\alpha d_{3}(\partial_{1}D_{0})^4\right]+\nonumber\\
&&\frac{1}{r^{2}}\left[d_{2}(\partial_{2}D_{0})(\partial_{3}D_{0})+\alpha d_{2}
(\partial_{2}D_{0}){(\partial_{3}D_{0})^3}-d_{3}(\partial_{2}D_{0})^{2}-\alpha d_{3}(\partial_{2}D_{0})^4\right]-m^2 d_{3}=0.\label{ab}
\end{eqnarray}
We take the phenomenon of variable split as
\begin{equation}
D_{0}=-\tilde{E}t+W(r,\theta)+J\phi,\label{c1}
\end{equation}
with $\tilde{E}=(E-J\Omega)$, $J$ and $E$ represent the energy at angle $\phi$ and particle's angular momentum, respectively. To consider a $4\times 4$ matrix, we apply Eq. (\ref{c1}) to Eqs. (\ref{aa})-(\ref{ab}) as
\begin{equation*}
K(d_{0},d_{1},d_{2},d_{3})^{T}=0.
\end{equation*}
The made matrix looks to be nontrivial. The components of it appear below:
\begin{eqnarray}
K_{00}&=&-W_{r}^2-\alpha W_{r}^4-\frac{1}{F(r)r^{2}}(W_{\theta}+\alpha W_{\theta}^4)
-\frac{1}{F(r)r^{2}\sin^2\theta}(\acute{J}^2+\alpha\acute{J}^4)-\frac{1}{F(r)}m^2,\nonumber\\
K_{01}&=&-(\tilde{E}+\alpha \tilde{E}^3)W_{r},\nonumber\\
K_{02}&=&-\frac{1}{F(r)r^{2}}(\tilde{E}+\alpha \tilde{E}^3)W_{\theta},\nonumber\\
K_{03}&=&-\frac{1}{F(r)r^{2}}(\tilde{E}+\alpha \tilde{E}^3)\acute{J},\nonumber\\
K_{10}&=&\tilde{E}W_{r}+\alpha \tilde{E}^3 W_{r}^4,\nonumber\\
K_{11}&=&\tilde{E}^2+\alpha \tilde{E}^4-
\frac{F(r)}{r^{2}}(W_{\theta}^2+\tilde W_{\theta}^4)
-\frac{F(r)}{r^{2}\sin^2\theta}(\acute{J}^2+\alpha\acute{J}^4)-F(r)m^2,\nonumber\\
K_{12}&=&\frac{F(r)}{r^{2}}(W_{r}+\alpha W_{r}^3)W_{\theta},~~
K_{13}=\frac{F(r)}{r^{2}\sin^2\theta}(W_{r}+\alpha W_{r}^3)W_{\theta},\nonumber\\
K_{20}&=&-\frac{1}{F(r)r^{2}}(\tilde{E}W_{\theta}+\alpha \tilde{E}W_{\theta}^3)\nonumber\\
K_{21}&=&\frac{F(r)}{r^{2}}(W_{\theta}+\alpha W_{\theta}^3)W_{r},\nonumber\\
K_{22}&=&\frac{1}{F(r)r^{2}}(\tilde{E}^2+\alpha \tilde{E}^4)
-\frac{F(r)}{r^{2}}(W_{r}^2+\alpha W_{r}^4)-\frac{1}{r^{4}\sin^2\theta}(\acute{J}^2+\alpha\acute{J}^4)
-\frac{1}{r^{2}}m^2\nonumber\\
K_{23}&=&\frac{1}{r^{4}\sin^2\theta}(W_{r}+\alpha W_{r}^3)W_{\theta},\nonumber\\
K_{30}&=&\tilde{E}(\acute{J}+\alpha \acute{J}^3),\nonumber\\
K_{31}&=-&\frac{F(r)}{r^{2}\sin^2\theta}(\acute{J}+\alpha \acute{J}^3)W_{r},~~
K_{32}=\frac{1}{r^{4}\sin^2\theta}(\acute{J}+\alpha \acute{J}^3)W_{\theta},\nonumber\\
K_{33}&=&\tilde{E}^2+\alpha\tilde{E}^4+
\frac{F(r)}{r^{2}\sin^2\theta}(W_{r}^2+\alpha W_{r}^4)-\frac{1}{r^{4}\sin^2\theta}(W_{\theta}^2+\alpha W_{\theta}^4)-
\frac{1}{r^{2}\sin^2\theta}m^2,\nonumber
\end{eqnarray}
with $\tilde{E}=\partial_{t}D_{0}$, $W_{r}=\partial_{r}{D_{0}}$, $W_{\theta}=\partial_{\theta}{D_{0}}$ and $\acute{J}=\partial_{\phi}D_{0}$.
Assuming the $K$ determinant is non-trivial, set $K$ to zero and lead the imaginary component to operate as
\begin{equation}\label{a1}
ImW^{\pm}= \pm\int\sqrt{\frac{\tilde{E}^{2}
+R_{1}\left[1+\alpha\frac{R_{2}}{R_{1}}\right]}{F(r)r^{-2}\sin^{-2}\theta}^2}dr,
\end{equation}
with
\begin{equation}
R_{1}=-\frac{W_{\theta}}{r^{2}\sin^{2}\theta}-\frac{m^2}{r^{2}\sin^{2}\theta},~~~
R_{2}=\frac{F(r)}{r^{2}\sin^{2}\theta}W_{r}^4+\tilde{E}^{4}+\frac{W_{r}^4}{r^{2}\sin^{2}\theta}.\nonumber
\end{equation}
The Eq. (\ref{a1}) implies
\begin{equation}
Im W^{+}=\pi\frac{\tilde{E}}{2\textsc{K}(r_{+})}\left[1+\Xi\alpha\right],
\end{equation}
where $\textsc{K}(r)$ indicates BH surface gravity and $\Xi$ indicates the kinetic energy factor along the horizon surface's tangent plane at the radiation that was produced point. To explain tunneling radiation mathematically, pair creation is proposed. In a region with intense gravitational forces, the two particles in a pair may be split apart as they can simultaneously annihilate. By quantum physics, annihilation occurs when a positive particle appears as a tunnel outside (radiation) and another negative particle tunnels inside the BH.
The following expression can be used to find the improved tunneling rate ($\textsc{T}$) for boson particles
\begin{equation}
\textsc{T}(W^{+})=\exp\left[-4Im W^+\right]=
\exp\left[{-2\pi}\frac{\tilde{E}}
{\textsc{K}(r_{+})}\right]\left[1+\Xi\alpha\right].
\end{equation}
Black holes are areas of extreme gravitational attraction into which surrounding matter gets pulled by gravitation. Classically, gravitation is so strong that nothing, even electromagnetic radiation, can escape from the BH. Hawking defined a BH as a cosmic monster that produces radiation due to quantum phenomena. The particle (positive mass)-antiparticle (negative mass) pairs are accelerated towards the BH's event horizon by vacuum fluctuations, according to the physics explanation of this emission process. A particle with positive mass has the energy to escape the BH, whereas a particle with negative mass cannot and would, therefore, fall in the BH, Hawking realized. Eventually, this action of a negative energy particle dropping diminishes the BH's mass. Further, a particle that travels far away might be considered thermal radiation. This process is a quantum tunneling effect, where a pair of particles will originate from the vacuum, and one of them will tunnel outside the event horizon with a complex action that looks like Hawking radiation since it has positive energy. These radiations depend on the BH's charge, mass, and directional momentum. The Hawking temperature for Kiselev-like AdS spacetime in the context of $f(R,~T)$ gravity can be obtained by applying the Boltzmann factor $\textsc{T}_{W}=\exp\left[-\tilde{E}/T' _{H}\right]$ under the influence of the GUP parameter in this manner
\begin{equation}
T'_{H}=\frac{1}{4\pi}\left[\frac{{1}}{r_{+}}+\frac{3r_{+}}{l^{2}}+\frac{3k(\zeta-8\pi\omega-3\omega)r_{+}^{\frac{8\pi(2+3\omega)+\zeta(3+7\omega)}{-8\pi-3\zeta+\omega\zeta}}}{8\pi+3\zeta-\omega\zeta}\right]\left[1-\alpha\Xi\right].\label{F5}
\end{equation}
This shows that the corrected Hawking temperature depends on the spacetime geometry and the quantum corrections. The original Hawking, semi-classical, and zero-order correction terms are the same. However, the first-order correction term needs to be smaller than the previous term and still meet GUP. The $T'_{H}$ is dependent on GUP ($\alpha$) parameter, $\omega$ parameter, the $f(R,~T)$ gravity $(\zeta)$ parameter, and cosmological constant. Furthermore, by excluding the gravity parameter $\alpha=0$, we can get the original temperature of the Kiselev-like AdS spacetime in the context of $f(R,~T)$ gravity. When $\alpha=0=\zeta$, we get Hawking temperature Kiselev-like AdS spacetime. The Schwarzschild spacetime Hawking temperature is the first term of our Hawking temperature result. 

The usual mechanisms of particle tunneling can be studied via the reduction of dimensions at the horizon \cite{F29, F30}. The Rindler space is essentially what all massive, non-extremal BHs resemble. All species of particles at the horizon for the standard Hawing radiation are virtually mass-less when considering infinite blueshift so that all particle Hawing temperatures are comparable. Our calculations also demonstrate that particles with various identities or quantum numbers should have varied effective Hawking temperatures due to quantum gravity effects. The existence of minimal length keeps the particles from ever being infinitely blueshifted as they get closer to the horizon.

The $T'_{H}$ of a Kiselev-like AdS spacetime in the context of $f(R,~T)$ gravity is irregular due to the quantum gravity impact, as $\Xi$ indicates the function of $\theta$ such that
\begin{equation}
\Xi=6\left(\frac{J^{2}_{\phi}\csc^{2}\theta+J^{2}_{\theta}}{r^{2}_{+}}+m^{2}\right)>0 .\label{c1}  
\end{equation}
By inserting Eq. (\ref{c1}) into Eq. (\ref{F5}), we get the corrected temperature as
\begin{equation}
T'_{H}=\frac{1}{4\pi}\left[\frac{{1}}{r_{+}}+\frac{3r_{+}}{l^{2}}+\frac{3k(\zeta-8\pi\omega-3\omega)r_{+}^{\frac{8\pi(2+3\omega)+\zeta(3+7\omega)}{-8\pi-3\zeta+\omega\zeta}}}{8\pi+3\zeta-\omega\zeta}\right]\left[1-6\alpha\left(\frac{J^{2}_{\phi}\csc^{2}\theta+J^{2}_{\theta}}{r^{2}_{+}}+m^{2}\right)\right].\label{F6}
\end{equation}
The $T'_{H}$ is related to the BH's surrounding fields like phantom, quintessence, radiation, and dust fields.
The $\omega=-\frac{4}{3}$ (phantom field), $\omega = -\frac{2}{3}$ (quintessence field), $\omega= \frac{1}{3}$ (radiation field), and $\omega=0$ (dust field) are among the specific circumstances that influence $T'_{H}$. Additionally, the third law of thermodynamics is violated by the negative temperature. Although negative temperature inside BHs has been studied previously \cite{F31, F32, F33}.
Furthermore, quantum techniques explicitly oppose temperature rise during evaporation, gradually equilibrating it completely. Naturally, there will be remnants of BHs.

Comparing the BH tunneling strategy to other temperature computation techniques exposes a great deal. Among these features, the tunneling strategy offers a dynamic model of the BH radiation, making it particularly intriguing for determining the BH temperature. The plan is beneficial when not wanting to include back-reaction effects in explaining the BH evaporation procedure. While the different approaches to calculating the thermodynamical features of BHs have been viable, they are unsatisfactory because they ignore the dynamic nature of the radiation and the procedures. After all, the fundamental geometry is fixed in most of these cases, including Hawking's case. The tunneling strategy's appropriate use in analyzing the effects of quantum gravity on BH radiation is one of its unique and intriguing features. According to the tunneling formalism, one of the particles can tunnel quantically through the BH horizon when the particle pairs are produced inside the event horizon. It is more reasonable to assume that some details on the BH's internal structure, such as the effects of quantum gravity, can be expressed in its radiation and made available to external observers.

These graphs depict the corrected temperature \(T'_H \) as a function of the horizon radius \(r_+ \) for different values of the parameter \(\alpha \), with fixed parameters \(\zeta \), $k$ and $l$ under the effects of dust ($\omega=0$) and radiation field ($\omega=1/3$). A comprehensive examination of each graph emphasizing the parameters' effect is below.
\begin{figure}[H]
\centering
\includegraphics[width=6cm,height=6cm]{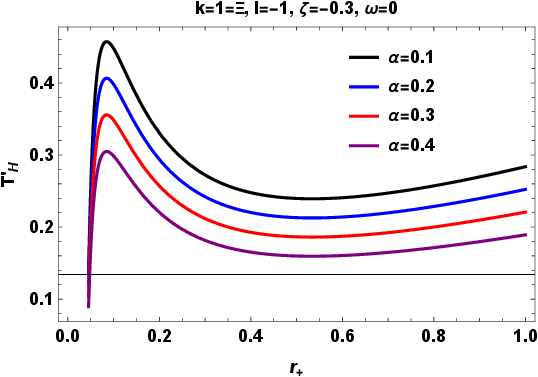}\includegraphics[width=6cm,height=6cm]{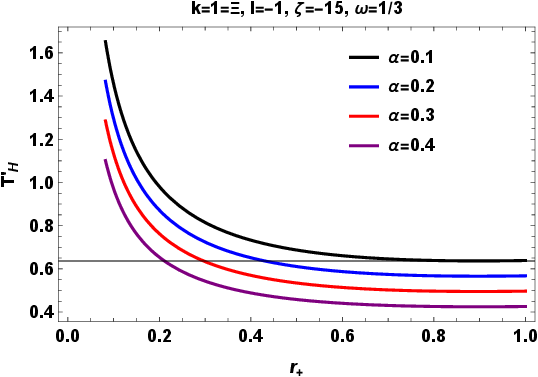}
\caption{Corrected temperature $T'_H$ via horizon radius $r_+$ for dust field ($\omega=0$) and radiation field ($\omega=1/3$) under the influence of quantum gravity parameter $(\alpha)$.}\label{te1}
\end{figure}

In the left plot of Fig. \ref{te1}, the parameters are set as \( k = 1 \), \( l = -1 \), \( \zeta = -0.3 \), and \( \omega = 0 \). As \( r_+ \) rises, the temperature \( T'_H \) falls monotonically, asymptotically reaching zero for every curve. This implies that the temperature stabilizes and approaches a lower bound as the horizon radius increases. The curves move downward as \( \alpha \) rises, resulting in lower temperatures for the same \( r_+ \). This suggests that the stability of the BH solution is impacted by a cooling effect with larger \( \alpha \) values. The graphs may show a horizon stability threshold, indicating a minimal value of \( r_+ \) below which \( T'_H \) becomes undefined or less physical.

In the right plot of Fig. \ref{te1}, the parameters change to \( k = 1 \), \( l = -1 \), \( \zeta =-15 \), and \( \omega = 1/3 \).
In contrast to the graph on the left, the temperature \( T'_H \) peaks at a specific \( r_+ \) before asymptotically declining. This peak suggests a thermodynamic phase transition in which the BH may change states. As in the graph on the left, the temperature decreases as \( \alpha \) increases. However, every \( \alpha \) value has a unique peak, suggesting a more intricate thermodynamic structure and maybe many phases or instabilities. The presence of a peak and subsequent decline in \( T'_H \) indicates that for each \( \alpha \), there exists an optimal horizon radius for maximum temperature, beyond which the system cools as \( r_+ \) increases.

From Fig. \ref{te1}, we conclude that the temperature profile changes from a monotonic drop to a peak and fall behavior when the left graph's values of \( \zeta = -0.3 \) and \( \omega = 0 \) are replaced with \( \zeta = -15 \) and \( \omega = 1/3 \) in the right graph. This suggests that the thermodynamic characteristics of the BH are sensitive to these factors. The left graph shows a stable, monotonically cooling system as the radius grows, while the peak on the right graph suggests possible instabilities or phase transitions. These graphs demonstrate how the horizon radius and factors like \( \alpha \), \( \zeta \), and \( \omega \) affect temperature, highlighting the intricate thermodynamics of the system. Understanding fundamental BH physics requires a knowledge of the different stability areas and potential phase transitions suggested by the variances in behavior.

\begin{figure}[H]
\centering
\includegraphics[width=6cm,height=6cm]{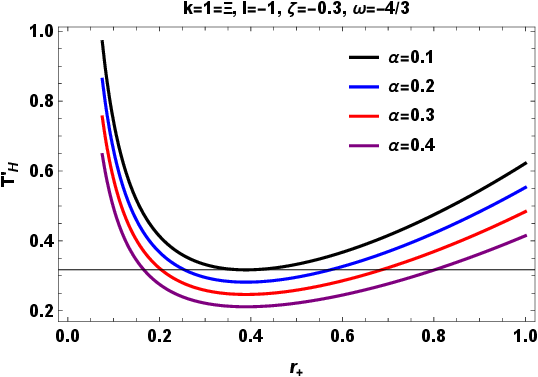}\includegraphics[width=6cm,height=6cm]{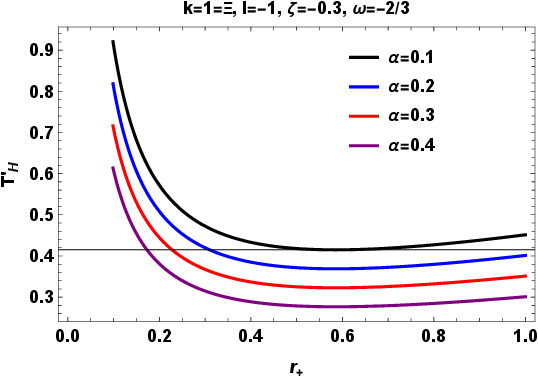}
\caption{Corrected temperature $T'_H$ via horizon radius $r_+$ for phantom field ($\omega=-4/3$) and quintessence field ($\omega=-2/3$) under the influence of quantum gravity parameter $(\alpha)$.}\label{t2}
\end{figure}

The left and right plots in Fig. \ref{t2} show the corrected temperature \(T'_H \) as a function of horizon radius \(r_+ \) for different values of the parameter \(\alpha \) in a BH context, together with additional parameters like \(k, l, \zeta \), and \(\omega \) for phantom field ($\omega=-4/3$) and quintessence field ($\omega=-2/3$). A critical analysis examines these parameter changes' patterns, shifts, and effects.

Increasing \(\alpha \) in both figures causes the curves to drop downward, showing that the corrected temperature \(T'_H \) at a given horizon radius decreases. The curves show a minimum point, which indicates the lowest feasible temperature before increasing with \(r_+ \). This may indicate a phase shift or a change in stability. The left graph, with \(\omega = -4/3 \), suggests that the temperature climbs more sharply after hitting the lowest than the right graph (\(\omega = -2/3 \)). A lower \(\omega \) value leads to lower temperatures across all values of \(r_+ \), demonstrating that \(\omega \) has a cooling function in this model.
Increasing \(\alpha \) consistently reduces the temperature profile, potentially due to its function in changing the BH's thermal characteristics or horizon behavior. For more significant \(\alpha \) (e.g., $0.4$), the temperature remains relatively low even as \(r_+ \) grows, demonstrating a dampening effect on temperature growth with horizon size. A minimum in \(T'_H \) indicates a critical moment where BH stability or phase properties may vary. As \(\alpha \) and \(\omega \) change, these minima move, presumably showing how changes in these parameters alter the stability regime of the BH.

From Fig. \ref{t2}, we conclude that the graphs show that both \(\alpha \) and \(\omega \) have a significant impact on temperature dynamics. Increasing \(\alpha \) and reducing \(\omega \) result in lower temperature profiles, indicating their influence on BH thermal properties and possible stability conditions at varying horizon radii.
   
    \begin{table} [h!]
    \centering
     \begin{tabular} { |p{4.5cm}|p{1.2cm}|p{1.2cm}|p{1.2cm}|p{1.2cm}|p{1.2cm}|}
    \hline
     $\omega$ & $r_+$ & $\zeta$ & $\alpha$ & $T_{H}$ & $T'_{H}$\\ [1.5 ex]
    \hline        Dust field ($\omega=0$) & 0.2 & 0.30 & 0.1 & 0.511 & 0.419\\
        \hline
        Radiation field ($\omega=\frac{1}{3}$) & 0.4 & 0.35 & 0.2 & -0.956 & -860 \\
        \hline
       Phantom field ($\omega=\frac{-4}{3}$) & 0.6 & 0.40 & 0.3 & 0.399 & 359\\
       \hline
        Quintessence field ($\omega=\frac{-2}{3}$) & 0.8& 0.45  & 0.4 & 0.461 & 0.415\\
    \hline
    \end{tabular}
    \caption{The $T_{H}$ and $T'_{H}$ are observed at different fields, the $f(R,~T)$ gravity $(\zeta)$ parameter and quantum gravity ($\alpha$) parameter, but the $l=1$, $k=1$, and $\Xi=1$ are fixed. The event horizons are different locations; we consider only the outer horizon, like $0.2\leq r_{+}\leq 0.8$.}
    \end{table}   
In the dust, radiation, phantom, and quintessence fields, the Hawking temperature is defined at $0.2\leq r_{+}\leq 0.8$. Further, as shown in Table I, the radiation field at both temperatures behaves negatively; this negative behavior evaluates the third law of thermodynamics, and the other positive evaluates the first law of thermodynamics. Since the relative cosmological constant, integration constant, and mass are fixed, we can see that the standard temperature is higher than the corrected temperature. We used the data in Table I to examine the quantum gravity reflect to reduce the temperature. In the BH thermodynamics, negative temperature indicates a transition between different thermodynamic phases, which are frequently connected with quantum corrections or modified gravity effects. It denotes a state of thermodynamic instability in which phenomena like negative specific heat and entropy corrections become relevant. This behavior is critical for understanding the interaction between BH evaporation, entropy corrections, and quantum gravitational effects, which provides knowledge about BH's complicated thermodynamic nature.

\section{Entropy Corrections for Kiselev-like AdS BHs within $f(R, T)$ gravity}

The Bekenstein-Hawking entropy connected to the correction factor in the theory of quantum loop expansion is examined in this section. The standard entropy of BH has an extra loop representing the logarithmic expression. Primordial BHs are very dependent on the entropy of BH \cite{E11}. It created a minimum mass for primordial BHs, allowing minor BHs to evaporate in their existing form. 
We examine entropy modifications for Kiselev-like AdS BHs in the context of $f(R, T)$ gravity. Banerjee and Majhi \cite{E12, E13, E14} have researched the entropy adjustments via the null geodesic approach. In the dynamic study of the new Schwarzschild BH with logarithmic corrections \cite{E14a} and their intense gravity, BHs never have a smaller area since they consume everything nearby. This leads to the idea of logarithmic adjustments, which were necessary corrections to Bekenstein's area-entropy relation. To achieve the new Schwarzschild BH corrected entropy, thermal fluctuations affect the stability of BHs with small radii, such as the new Schwarzschild BH; as a result, these first-order corrections cause unstable zones to form around small BHs.
In the framework of $f(R, T)$ gravity, we examine the entropy corrections for Kiselev-like AdS BHs using the Bekenstein-Hawking entropy formula \cite{E15} for $1^{st}$ order corrections.
In the setting of $f(R, T)$ gravity, we compute the logarithmic entropy corrections for Kiselev-like AdS BHs using the standard entropy $\mathbb{S}_{T}$ and the formulations of $T'_{H}$ given as follows
\begin{equation}
\mathbb{S}_{f}=\mathbb{S}_{f_{0}}-\frac{1}{2}\ln\Big|T'^{2}_{H} \mathbb{S}_{f_{0}}\Big|+...~.\label{vv}
\end{equation}
It has been discovered that the entropy of the spacetime is always equivalent to a quarter of the BH horizon's area. Applying the approach to tunneling provides the finest means of imagining the radiation source. The phenomenon of particles being able to pass through energy barriers inside electronic systems is known as quantum tunneling. An electric field is necessary for forming electron-positron pairs, which is the framework of this strategy. The following formula may be used to determine the standard entropy for Kiselev-like AdS BHs in the setting of $f(R, T)$ gravity
\begin{equation}
\mathbb{S}_{f_{0}}=\frac{A_f}{4},\label{af}
\end{equation}
where
\begin{equation}
A_f=\int_{0}^{2\pi}\int_{0}^{\pi}\sqrt{g_{\theta\theta}g_{\phi\phi}}d\theta d\phi=4\pi r_+^2.\label{af1}
\end{equation}
In the context of $f(R, T)$ gravity, the standard entropy for Kiselev-like AdS BHs can be computed by putting the value of $A_f$ from Eq. (\ref{af1}) into Eq. (\ref{af}) as follows 
\begin{equation}
\mathbb{S}_{f_0}=\pi r_+^2.\label{v1}
\end{equation}
After putting the values from (\ref{v1}) into Eq. (\ref{vv}), we obtain the entropy corrections as
\begin{eqnarray}
\mathbb{S}_{f}=\pi r_+^2
-\frac{1}{2}\ln\left|\frac{r^2_{+}}{16\pi}\left[\frac{{1}}{r_{+}}+\frac{3r_{+}}{l^{2}}+\frac{3k(\zeta-8\pi\omega-3\omega)r_{+}^{\frac{3(3\omega\zeta+8\pi\omega-\zeta)}{8\pi+3\zeta-\omega\zeta}}}{8\pi+3\zeta-\omega\zeta}\right]^2\left[1-\alpha\Xi\right]^2.\right|+...,\label{b2}
\end{eqnarray}

The Eq. (\ref{b2}) shows the corrected entropy for Kiselev-like AdS BHs in the context of $f(R,~T)$ gravity. The corrected entropy $S_{f}$ is dependent on GUP ($\alpha$) parameter, $\omega$ parameter, the $f(R,~T)$ gravity $(\zeta)$ parameter, cosmological constant. Furthermore, by excluding the gravity parameter $\alpha=0$, we can get the original entropy without quantum effects of the Kiselev-like AdS spacetime in the context of $f(R,~T)$ gravity. When $\alpha=0=\zeta$, we get entropy for Kiselev-like AdS spacetime. The Schwarzschild spacetime entropy is the first term of our entropy result. 

The graphical analysis of entropy corrections $S_{f}$ versus horizon $r_+$ for  Kiselev-like AdS BHs have been analyzed in this section. We analyze the impacts of $\alpha$ and $\omega$ on corrected entropy $S_{f}$ via graphs.
\begin{figure}[H]
\centering
\includegraphics[width=6cm,height=6cm]{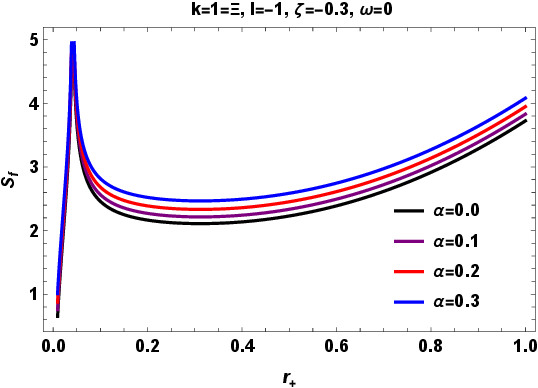}\includegraphics[width=6cm,height=6cm]{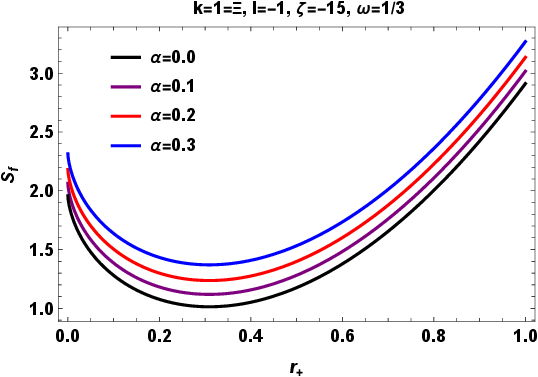}
\caption{Corrected entropy $S_f$ via horizon radius $r_+$ for dust field ($\omega=0$) and radiation field ($\omega=1/3$) under the influence of quantum gravity parameter $(\alpha)$.}\label{e1}
\end{figure}
The Fig. \ref{e1} exhibits the corrected entropy $S_f$ via horizon radius $r_+$ for various values of $\alpha=0,~0.1,~0.2,~0.3.$. In the left plot, the entropy diverges dramatically near $r_+\rightarrow 0$, as it usually does when quantum corrections dominate owing to short horizon lengths. For $r_+\approx 0.2$, the entropy stabilizes and progressively increases. The $\alpha$ causes a greater entropy at $r_+$ compared to $\alpha=0$ (without quantum effects), indicating a positive correction from the dust field. 
For $r_+\approx 1$, entropy increases more strongly with higher $\alpha$. As $\alpha$ grows, the curve gap widens, indicating a more substantial impact of the dust field at broader horizon scales.

Compared to the dust scenario, the divergence around $r_+\rightarrow 0$ in the right plot is less apparent. This suggests that quantum corrections under the radiation field have a more negligible influence on tiny scales. Entropy approaches a minimum at an intermediate $r_+$, which is significantly lower for larger values of $\alpha$. This distinct behavior might be attributable to the interaction of radiation effects and quantum correction. Entropy rises gradually for $r_+>0.6$, with greater $\alpha$ trending towards higher entropy levels. The curves stay almost parallel, suggesting that $\alpha$ has a constant effect over wide radii.

From Fig. \ref{e1}, we conclude that in contrast to the radiation field $(\omega=1/3$), the dust field $(\omega=0$) shows a more significant divergence, indicating more substantial quantum effects in the former. The existence of minima in the radiation field highlights a unique thermodynamic behavior not seen under the dust field. This could suggest a fluctuating phase influenced by radiation. While raising $\alpha$ systematically raises $S_f$ for both fields, the dust field accelerates the rise speed.
Dust fields amplify more entropy corrections than radiation fields, especially for more extraordinary horizon lengths. This sheds light on how various cosmological fields affect BH thermodynamics and may give a starting point for research on quantum gravity corrections in the presence of different external factors.

\begin{figure}[H]
\centering
\includegraphics[width=6cm,height=6cm]{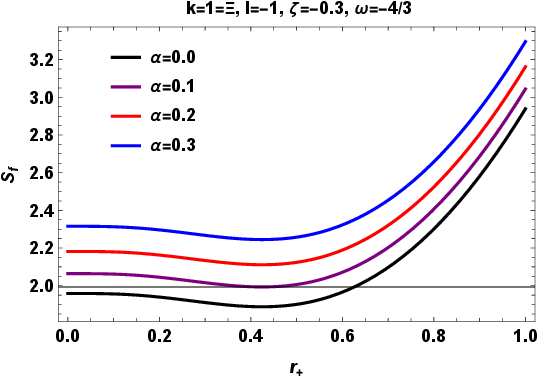}\includegraphics[width=6cm,height=6cm]{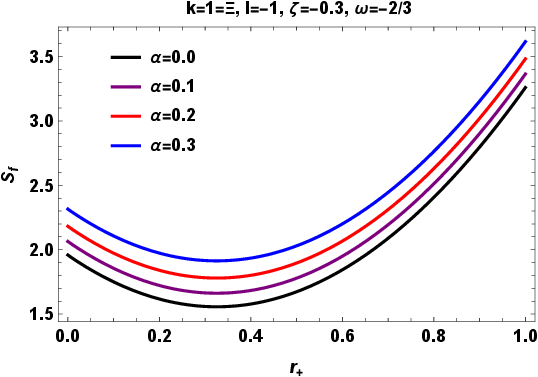}
\caption{Corrected entropy $S_f$ via horizon radius $r_+$ for phantom field ($\omega=-4/3$) and quintessence field ($\omega=-2/3$) under the influence of quantum gravity parameter $(\alpha)$.}\label{e2}
\end{figure}
Figure \ref{e2} shows how a phantom field and a quintessence field affect the corrected entropy $S_f$ behavior as a function of the horizon radius, $r_+$. The phantom field ($\omega=-4/3$) and quintessence field ($\omega=-2/3$) have different values of the parameter $\omega$, which distinguishes these fields. The entropy correction appears to be affected by the parameter $\alpha$, which fluctuates in increments of $0.1$ from $0.0$ to $0.3$ over a range of $r_+$ values.

In the left plot of Fig. \ref{e2}, with increasing $r_+$, the modified entropy $S_f$ first falls and then rises, creating a minimum. This behavior implies that the entropy adjustment is not linear for the horizon size in the presence of a phantom field.
The entropy correction curve rises higher as $\alpha$ increases. This suggests that the parameter $\alpha$ amplifies the corrected entropy in the presence of a phantom field as higher values of $\alpha$ increase the total entropy at any fixed $r_+$.
Every curve has a unique minimum, and as $\alpha$ varies, so do its position and depth. This minimum may indicate a balance between the internal and external components of the system since it implies an equilibrium horizon size where the phantom field least impacts the entropy correction.

In the right plot of Fig. \ref{e2}, the modified entropy $S_f$ steadily falls reaches a minimum, and rises more abruptly with higher $r_+$, in contrast to the phantom field example. The curve's more noticeable U-shape suggests that quintessence fields affect entropy differently than phantom fields. As in the previous plot, the adjusted entropy increases for all values of $r_+$ as $\alpha$ increases. However, the upward shift caused by $\alpha$ seems more substantial, indicating that the quintessence field significantly affects entropy correction more than the phantom field. The curves show a sharper rise in $S_f$ after reaching the minimum compared to the phantom field example, but the minimum position is still controlled by $\alpha$.

From Fig. \ref{e2}, we conclude that the effect of the quintessence field creates a sharper U-shape in $S_f$, whereas that of the phantom field is more gradual, indicating that quintessence fields contribute more to entropy at greater horizon radii. The parameter $\alpha$ consistently increases the entropy in both areas, indicating a positive link. The impact is more substantial in the quintessence field, suggesting that $\alpha$ has a field-dependent effect on entropy.
The minimum points in each figure indicate an equilibrium horizon size where the entropy correction from the horizon field effects is at its lowest. This characteristic may indicate stable BH structures when phantom and quintessence fields are present.

\section{Discussion}
Parikh and Wilczek suggested a technique to obtain Hawking radiation that utilizes quantum tunneling. We consider that a particle-antiparticle pair forms near the horizon in a BH. Particles' associated vector fields can be subdivided into ingoing modes that aim to move inside the BH horizon and outgoing forms that radiate it. The horizon contains the imprisoned ingoing boson particles. However, with the horizon acting as a barrier to some extent when the outgoing boson particles radiate the BH, some outgoing modes are analyzed to the quantum tunneling phenomenon. We can consider the particle outside the horizon to be the BHs radiation if the particle that exits our universe has positive energy. Such a particle, as well as an antiparticle, can exist continuously. We used classically prohibited paths to compute the particle's WKB approximate for tunneling probability. It determined the Hawking temperature by applying semi-classical to compare the probability with the Boltzmann factor.

 It's crucial to study that we have disregarded the effects of self-gravity and particle radiation back-reaction. We have modified the Lagrangian equation that characterizes the spin-$1$ particle's motion. Then, assuming the Hamilton-Jacobi approach, we find the tunneling probabilities and effective temperatures of particles from the given BHs. In addition to the BH's properties, the corrected tunneling depends on the released particles' energy, mass, and angular momentum. 

 The corrected Hawking temperature for Kiselev-like AdS BH depends on the spacetime geometry and the quantum corrections. The original Hawking, semi-classical, and zero-order correction terms are the same. However, the first-order correction term needs to be smaller than the previous term and still meet GUP. The corrected temperature $T'_{H}$ is dependent on GUP ($\alpha$) parameter, $\omega$ parameter, the $f(R,~T)$ gravity $(\zeta)$ parameter, and cosmological constant. Furthermore, by excluding the gravity parameter $\alpha=0$, we can get the original temperature of the Kiselev-like AdS spacetime in the context of $f(R,~T)$ gravity. When $\alpha=0=\zeta$, we get Hawking temperature Kiselev-like AdS spacetime. The Schwarzschild spacetime Hawking temperature is the first term of our Hawking temperature result. The $T'_{H}$ is related to the BH's surrounding fields like phantom, quintessence, radiation, and dust fields.
The $\omega=-\frac{4}{3}$ (phantom field), $\omega = -\frac{2}{3}$ (quintessence field), $\omega= \frac{1}{3}$ (radiation field), and $\omega=0$ (dust field) are among the specific circumstances that influence $T'_{H}$. Moreover, we graphically analyzed the behavior of standard temperature, corrected temperature, and modified entropy under the effects of different fields. We have graphically examined the impacts of $f(R,~T)$ gravity parameter $\zeta$, GUP parameter $\alpha$ in the presence of dust, radiation, phantom, and quintessence field.
The graphical results are summarized as follows:

\textbf{Impacts of Dust and Radiation Fields}
\begin{itemize}
 \item \textbf{Standard temperature}:  The thermal response is always influenced by the parameter \( \zeta \), and its value determines whether the temperature rises or falls. The connection between peak temperature and \( \zeta \) is inverse. A physical system where $\zeta$ may represent an external field or interaction strength that affects the thermal equilibrium states is probably modeled by the temperature behavior dependent on $r_+$ and $\zeta$.
 
 \item \textbf{Corrected temperature}: The temperature profile changes from a monotonic drop to a peak and fall behavior when the values of \( \zeta = -0.3 \) and \( \omega = 0 \) are replaced with \( \zeta = -15 \) and \( \omega = 1/3 \). This suggests that the thermodynamic characteristics of the BH are sensitive to these factors. The graph shows a stable, monotonically cooling system for the dust field as the radius grows, while the peak on the radiation field suggests possible instabilities or phase transitions. The graphs demonstrate how the horizon radius and factors like \( \alpha \), \( \zeta \), and \( \omega \) affect temperature, highlighting the intricate thermodynamics of the system.

 \item \textbf{Corrected entropy}: In contrast to the radiation field $(\omega=1/3$), the dust field $(\omega=0$) shows a more significant divergence, indicating more substantial quantum effects in the former. The existence of minima in the radiation field highlights a unique thermodynamic behavior not seen under the dust field. This could suggest a fluctuating phase influenced by radiation. While raising $\alpha$ systematically raises $S_f$ for both fields, the dust field accelerates the rise speed. Dust fields amplify more entropy corrections than radiation fields, especially for more extraordinary horizon lengths.
 \end{itemize}

\textbf{Impacts of Phantom and Quintessence Fields}

\begin{itemize}
 \item \textbf{Standard temperature}:  Increasing \( \zeta \) often results in greater temperatures at bigger \( r_+ \) values for both values of \( \omega \), indicating that \( \zeta \) affects the system's thermal stability or rate of temperature growth. The behavior of the temperature at low \( r_+ \) implies that the system is more sensitive to parameter changes for tiny BHs or horizons, which may indicate phase transitions or critical events related to horizon formation.
 
\item \textbf{Corrected temperature}: the graphs show that both \(\alpha \) and \(\omega \) have a significant impact on temperature dynamics. Increasing \(\alpha \) and reducing \(\omega \) result in lower temperature profiles, indicating their influence on BH thermal properties and possible stability conditions at varying horizon radii.

\item \textbf{Corrected entropy}: the effect of the quintessence field creates a sharper U-shape in $S_f$, whereas that of the phantom field is more gradual, indicating that quintessence fields contribute more to entropy at greater horizon radii. The parameter $\alpha$ consistently increases the entropy in both areas, indicating a positive link. The impact is more substantial in the quintessence field, suggesting that $\alpha$ has a field-dependent effect on entropy. The minimum points in each figure indicate an equilibrium horizon size where the entropy correction from the horizon field effects is at its lowest. This characteristic may indicate stable BH structures when phantom and quintessence fields are present.
\end{itemize}

\section*{Acknowledgement}

The paper was funded by the National Natural Science Foundation of China 11975145.

\section*{Conflict of Interest}
The authors declare no conflict of interest.

\section*{Data Availability Statement}
The data supporting this study's findings are available from the corresponding author upon reasonable request.

\end{document}